\documentclass[12pt]{JHEP3}
\pdfoutput=1
\usepackage{amsfonts, amsmath, amsthm}
\usepackage{graphicx}

\def\f{\tilde{f}}

\def \Tr {\mathop{\rm Tr}\nolimits}

\newcommand\ignore[1]{}
\def\one{{\,\hbox{1\kern-.8mm l}}}

\def\a{\alpha}\def\b{\beta}

\def\Dslash{D\!\!\!\!/\,\,}

\newcommand{\Cset}{{\,\,{{{^{_{\pmb{\mid}}}}\kern-.45em{\mathrm C}}}}}

\def \d {\partial}
\newcommand{\be}{\begin{equation}}
\newcommand{\ba}{\begin{eqnarray}}
\newcommand{\bea}{\begin{eqnarray}}

\newcommand{\ee}{\end{equation}}
\newcommand{\eea}{\end{eqnarray}}
\newcommand{\ea}{\end{eqnarray}}

\title{On abelianizations of the ABJM model and applications to condensed matter}
\author{Jeff Murugan$^{1}$\footnote{jeff@nassp.uct.ac.za} and Horatiu Nastase$^{2}$
\footnote{nastase@ift.unesp.br} \\

$^{1}$The Laboratory for Quantum Gravity \& Strings, \\
Department of Mathematics and Applied Mathematics, \\
University of Cape Town, \\
Private Bag, Rondebosch, 7700, \\
South Africa.\\

$^{2}$Instituto de F\'{i}sica Te\'{o}rica,\\ UNESP-Universidade Estadual Paulista,\\
R. Dr. Bento T. Ferraz 271, Bl. II,\\ S\~ao Paulo 01140-070, SP, Brazil}

\abstract{In applications of AdS/CFT to condensed matter systems in 2+1 dimensions, the ABJM model is often used, however the condensed matter models 
are usually abelian and contain charged fields. We show that a naive reduction of the ABJM model to $N=1$ does not have the desired features, but we 
can find an abelian reduction that has most features, and we can also add fundamental fields to the ABJM model to obtain other models with similar properties.
} 

 \preprint{QGaSLAB-13-02}

\begin{document}

\maketitle

\section{Introduction}

The last few years has seen an overwhelming amount of work devoted to applying the AdS/CFT correspondence to condensed matter systems (see \cite{Hartnoll:2009sz,Herzog:2009xv} for a 
review). The usual approach to this problem is somewhat 
phenomenological, since there doesn't exist any good way to derive an explicit AdS/CFT duality that gives a condensed matter system of interest. One obstacle for this is the necessity to use a large $N$ gauge theory in order to have a pure gravity dual (as opposed to a full string theory dual), 
whereas condensed matter models of interest are usually {\it abelian}. Another is that it appears to be very difficult to obtain relevant condensed matter models from a full string theory construction.

In two previous papers (together with A.Mohammed) \cite{Mohammed:2012rd,Mohammed:2012gi}, we have begun to address these questions. We have shown, in particular, that 
a bosonic abelian reduction of the ABJM model (which is considered a primer for 2+1 dimensional condensed matter systems in the same way that ${\cal N}=4$ SYM in 3+1
dimensions is considered a primer for QCD) reproduces a relativistic Landau-Ginzburg model which has been used for describing the quantum 
critical phase near a superconductor-insulator transition \cite{Myers:2010pk,Sachdev:2011wg}.
We have also shown that this is a nontrivial abelian truncation, still allowing for a gravity dual description, and that the abelianization procedure can be 
mimicked in a simple condensed matter model, at least as far as its general features.

In this paper we continue this program, and ask if it is possible to construct abelianizations of the ABJM model that result in actions with the same 
general features as condensed matter models with scalars, fermions and gauge fields, and both global and local charges. We will investigate the $N=1$ 
reduction of the ABJM model, as well as a generalization of the abelianization in \cite{Mohammed:2012rd,Mohammed:2012gi} that includes fermions, and the
introduction of fundamental fields in the ABJM model.\footnote{After the first version of this paper appeared on the arXiv, the paper \cite{Hyun:2013sf}
appeared, where however a different ansatz for the fermions is considered (with an extra $\epsilon^{\a\b}$), which leads to an ${\cal N}=4$ supersymmetric 
model (an additional ${\cal N}=2$ supersymmetric model uses two sets of matrices). It is also stated there that the gravity dual corresponding to
the abelianization, viewed as a perturbation around a vacuum, is strongly coupled and cannot be trusted. However in our case, it is impossible to view a 
general solution of the abelian reduction as a small pertubation, as it can be easily checked, hence this is not a problem for our construction.} 
We will find that simply setting $N=1$ does {\it }not serve the desired purpose, but the other two procedures have features that are similar to the condensed matter model.

The paper is organized as follows. In section 2 we describe the features of the condensed matter models that we are after followed by an analysis of the $N=1$ case in section 3. Section 4 concerns itself with the generalization of the abelianization in \cite{Mohammed:2012rd,Mohammed:2012gi} to include fermions. In 
section 5 we add fundamental degrees of freedom to the ABJM model and compare to existing literature and in section 6 we conclude with some thoughts on future directions.

\section{Condensed matter models}

In order to facilitate the utility of the gauge/gravity correspondence to (planar) condensed matter physics, various models have recently been proposed that bear at least some resemblance to what is quickly becoming the canonical example of an $AdS_{4}/CFT_{3}$ relation, the ABJM model \cite{Aharony:2008ug}. 
These models were then thought to 
be well described by the gravity dual to ABJM, namely type IIA superstring theory (or, at least, its supergravity limit) on $AdS_{4}\times \mathbb{CP}^{3}$. One characteristic they all share is  that they are {\it abelian} and, depending on the condensed matter application in question, contain an {\it emergent} (non-electromagnetic) $U(1)$ gauge field.

One such model, proposed in \cite{Huijse:2011hp}, which we will take as a relevant example, 
was used to describe compressible Fermi surfaces. It is this that we will take as a template to work with, since it exhibits several of the 
general features observed in most condensed matter systems modelled by the ABJM action. Defined through the nonrelativistic action
\bea
S&=&\int d^3x\left[f_+^\dagger\left((\d_\tau-iA_\tau)-\frac{(\vec{\nabla}-i\vec{A})^2}{2m_f}-\mu\right)f_+\right.\cr
&&\left. +f_-^\dagger\left((\d_\tau+iA_\tau)-\frac{(\vec{\nabla}+i\vec{A})^2}{2m_f}-\mu\right)f_-\right.\cr
&&\left.+b_+^\dagger\left((\d_\tau-iA_\tau)-\frac{(\vec{\nabla}-i\vec{A})^2}{2m_b}+\epsilon_1-\mu\right)b_+\right.\cr
&&\left.+b_-^\dagger\left((\d_\tau+iA_\tau)-\frac{(\vec{\nabla}+i\vec{A})^2}{2m_b}+\epsilon_1-\mu\right)b_-\right.\cr
&&\left.+\frac{u}{2}(b_+^\dagger b_++b_-^\dagger b_-)^2+vb_+^\dagger b_-^\dagger b_-b_+ -g_1(b_+^\dagger b_-^\dagger f_-f_+ +h.c.)\right.\cr
&&\left.+c^\dagger\left(\d_\tau-\frac{(\vec{\nabla})^2}{2m_c}+\epsilon_2-\mu\right)c-g_2(c^\dagger(f_+b_-+f_-b_+)+h.c.)\right],\label{hsmodel}
\eea
it has an emergent gauge field (i.e., not the electromagnetic gauge field), a $U(1)$
global symmetry current corresponding to electric charge, 
\be
Q=f_+^\dagger f_++f_-^\dagger f_-+b_+^\dagger b_++b_-^\dagger b_-+2c^\dagger c,\label{hselcharge}
\ee
and both fundamental charged bosons $b_\pm$ and fermions $\f_\pm$ coupled to the gauge field as well as a neutral fermion $c$. When writing a relativistic version of this, we should replace
\be
(\d_\tau-iqA_\tau)-\frac{(\vec{\nabla}-iq\vec{A})^2}{2m}-\mu\rightarrow 
(\d_\mu-iqA_\mu)^2-m^2,
\ee 
for bosons, and for fermions, by the square root of the Klein-Gordon operator above. 

We can now see the general features that we need for a good model: it needs to have bosons and fermions, coupled to an emergent abelian gauge field with both
positive and negative charges, and a global charge (corresponding to the electric charge) with values for the charges of the fields independent of the 
local charge. Moreover, we would like to find a model that has all the fields with positive global charges, but both positive and negative local charges.
The interaction terms will have up to fourth order couplings in the bosons, and two bose-two fermi interactions.

We will try to obtain models that have the same general properties using abelianization procedures for the ABJM model, and later by adding fundamental fields.

\section{ABJM $U(N)\times U(N)$ model for $N=1$}

The naive first thing to do in order to obtain an abelian field theory from the $U(N)\times U(N)$ ABJM model would be to set $N=1$. One might well be concerned about the sensibility of such an attempt but since it is the $U(N)\times U(N)$ version of the ABJM model that is related to a brane construction and {\it not} the $SU(N)\times SU(N)$ one, it is a consistent thing to try. Nevertheless, we demonstrate below that it is too naive.

The action was written down in \cite{Sasaki:2009ij,Nastase:2010ft}, as
\be\label{single}
S^{N=1}_{\mathrm{ABJM}}=\int d^3 x\left[\frac{k}{4\pi}\epsilon^{\mu\nu\lambda}\left(A_\mu^{(1)}\d_\nu A_\lambda^{(1)}-A_\mu^{(2)}\d_\nu A_\lambda^{(2)}\right) -i\psi^{A\dagger}\gamma^\mu D_\mu \psi_A-D_\mu C^\dagger_A D^\mu C^A\right]\;,
\ee
where the gauge covariant derivative
\be
D_\mu C^A=\d_\mu C^A+i(A_\mu^{(1)}-A^{(2)}_\mu)C^A\;.
\ee

In order to understand the physics of this truncation, we will need to Higgs the model. We know from general considerations \cite{Mukhi:2008ux,Chu:2010fk} 
that Higgsing the ABJM model means that the Chern-Simons gauge field will eat a scalar degree of freedom and become physical, i.e. of the Maxwell type. 
It is this behaviour that we want to reproduce here. 

The Higgsing procedure in a rather general case, for both $SU(N)\times SU(N)$ and $U(N)\times U(N)$ models was set out in \cite{Chu:2010fk}. To specialize to our case, we will follow this quite closely and simply set $N=1$. We begin by defining 
\bea
A_\mu^\pm&=&\frac{1}{2}(A_\mu^{(1)}\pm A_\mu^{(2)}),\cr
F_{\mu\nu}^\pm&=&\d_\mu A_\nu-\d_\nu A_\mu.
\eea
With this, we can easily check that the Chern-Simons part of the action becomes 
\be
\int d^3x\frac{k}{2\pi}\epsilon^{\mu\nu\lambda}A_\mu^-F_{\nu\lambda}^+.
\ee
In order to Higgs the model, it is necessary to write
\be
C^A=\frac{X^A+iX^{A+4}}{\sqrt{2}}+v\delta^{A4},
\ee
where the $X^A$ and $X^{A+4}$ are real, and we consider the scalar field VEV $v$ to be large. 
Then the covariant derivative of the scalars becomes
\be
D_\mu C^A=\d_\mu\left(\frac{X^A+iX^{A+4}}{\sqrt{2}}\right)+2iA_\mu^-\left(\frac{X^A+iX^{A+4}}{\sqrt{2}}+v\delta^{A4}\right),
\ee
while the kinetic term for the scalars reduces to
\bea
\left|D_\mu C^A\right|^2&=&\frac{(\d_\mu X^A-A_\mu^-X^{A+4})^2}{2}+\frac{(\d_\mu X^{A+4}+2A_\mu^-X^A+2\sqrt{2}vA_\mu^-\delta^{A4})^2}{2}\cr
&=&\frac{(\d_\mu X^A)^2+(\d_\mu X^{A+4}+2\sqrt{2}vA_\mu^-\delta^{A4})^2}{2},
\eea
where we have dropped terms subleading in $v$. On the other hand, the term
\be
\frac{k}{2\pi}\epsilon^{\mu\nu\lambda}\frac{1}{2\sqrt{2}v}\d_\mu X^8F_{\nu\lambda}^+
\ee
vanishes by partial integration and use of the Bianchi identity, so it can be added to the action with impunity. Consequently, the bosonic part of the action,
\bea
S_{\rm bose}=\int d^3x \left[\frac{k}{2\pi}\epsilon^{\mu\nu\lambda}\left(A_\mu^-+\frac{1}{2\sqrt{2}v}\d_\mu X^8\right)F^+_{\nu\lambda}
-\frac{(\d_\mu X^A)^2}{2}-\frac{(\d_\mu X^{A+4}+2\sqrt{2}vA_\mu^-\delta^{A4})^2}{2}\right].\nonumber
\eea
To clarify this, we now make the shift
\be
  A_\mu^-\rightarrow A_\mu^--\frac{1}{2\sqrt{2}v}\d_\mu X^8,\nonumber
\ee
after which the bosonic action becomes 
\be
  S_{\rm bose}=\int d^3x \left[\frac{k}{2\pi}\epsilon^{\mu\nu\lambda}A_\mu^-F_{\nu\lambda}^  
  +-\frac{(\d_\mu X^{I'})^2}{8}-4v^2(A_\mu^-)^2\right],
\ee
where $I'=1,...,7$ and again, to leading order in $v$. At this point, we recognize that the field $A_{\mu}^{-}$ is, in fact, auxilliary and can be integrated out to give
\be
  A_\mu^-=\frac{k}{16\pi v^2}\epsilon_{\mu\nu\lambda}F^{+\nu\lambda}.
\ee
Substituting this expression back in the bosonic action, we finally find for the Higgsed action 
\be
  S=\int d^3x \left[-\frac{k^2}{32\pi^2v^2}F^{+\mu\nu}F^+_{\mu\nu}-\frac{(\d_\mu X^{I'})^2}  
  {2}-i\bar\psi^{\dagger A}\gamma^\mu D_\mu \psi^A\right].
\ee
We see that we must define 
\be
\frac{k^2}{32\pi^2v^2}=\frac{1}{4g^2},
\ee
and therefore we must also have $k\rightarrow\infty$, such that the ratio $k/v$ remains fixed. This in turn means that, in the covariant derivative
\bea
  D_\mu \psi^A&\rightarrow& \left(\d_\mu +2i \left(A_\mu^--\frac{1}{2\sqrt{2}v}\d_\mu   
  X^8\right)\right)\psi^A
  =\left(\d_\mu +2i \left(\frac{1}{4g^2}\frac{2\pi}{k}\epsilon_{\mu\nu\lambda}F^{+\nu\lambda}  
  \right)\right)\psi^A,\nonumber
\eea
all terms additional to the partial derivative $\partial_{\mu}\psi^{A}$ are subleading in $v$ and can be dropped, giving the final form of the Higgsed action,  
\be
  S_{Higgs}=\int d^3x \left[-\frac{1}{4g^2}F^{+\mu\nu}F^+_{\mu\nu}-\frac{(\d_\mu   
  X^{I'})^2}{2}-i\bar\psi^{\dagger A}\gamma^\mu \d_\mu \psi^A\right].
\ee
In this form it is clear that, while we retain the the global $SO(7)$ symmetry (with $A$ a spinor index, and $I'$ a fundamental index),
the local symmetry does not act on the matter fields.
In other words, setting $N=1$ results in nothing more than a {\it free Maxwell supermultiplet}, with the scalars and fermions not coupled to the gauge field. 
This ``abelianization" therefore does {\it not} produce the desired features for applying to the condensed matter phenomena of interest to us.

\section{Nontrivial abelianization with fermions}

In a series of recent papers \cite{Mohammed:2012rd,Mohammed:2012gi}, we introduced a nontrivial abelianization procedure which, when further restricted, led to a 
relativistic Landau-Ginzburg model relevant for condensed matter. While promising, the focus there was strictly on the {\it bosonic} sector of the ABJM model. We now ask whether it is possible to extend this procedure to include fermions, and, moreover, whether 
the resulting effective theory (with scalars, gauge fields and fermions) exhibits any of the desired features of (\ref{hsmodel}).

In keeping with our previous notation, we split the scalars in the ABJM supermultiplet as $C^A=(Q^\a,R^{\dot\a})$, with the abelianization ansatz
\be
Q^\a=\phi_\a G^\a;\;\;\;\;
R^{\dot\a}=\chi_{\dot\a} G^{\dot\a}
\ee
with no sum over $\a, \dot\a=1,2$, and the $G^\a$ matrices as introduced in \cite{Gomis:2008vc}  (see also \cite{Terashima:2008sy}). Now we do the same for the fermions, splitting $\psi_A=(\psi_\a,\tilde \psi_{\dot\a})$. In analogy with the scalars, the abelianization ansatz for the fermions is 
\be
\psi_\a=\eta_\a G^\a;\;\;\;\; \tilde \psi_{\dot\a}=\tilde \eta_{\dot\a} G^{\dot\a}, 
\ee
again with no sum over $\a, \dot\a$. Then, by virtue of the fact that the bifundamental abelianization ansatz is the same as for the scalars, we can immediately show that the ABJM covariant derivative on fermions
reduces to the same {\it abelian covariant derivative} that acts on the abelian scalars, $D_\mu\eta_i=\left(\d_\mu-ia_\mu^{(i)}\right)\eta_i$, $D_\mu\tilde\eta_i=\left(\d_\mu-ia_\mu
^{(i)}\right)\tilde\eta_i$. It follows then that the kinetic term for the fermions is the standard Dirac one. It remains only to calculate the interaction terms between the scalars and the fermions. That is easily done, using the defining relation for the GRVV matrices $G^\a=G^\a G^\dagger_\b G^\a-G^\b G_\b^\dagger G^\a$. After some algebra, we find that the only non-vanishing term contributes,
\bea
&&\Tr\Big[C^\dagger_A C^A\psi^{B\dagger}\psi_B-\psi^{B\dagger}C^AC^\dagger_A \psi_B\Big]\cr
&&=\frac{N(N-1)}{2}\Big[(|\phi_1|^2+|\chi_1|^2)(\eta^\dagger_2\eta_2+\tilde\eta^\dagger_2\tilde\eta_2)
+(|\phi_2|^2+|\chi_2|^2)(\eta^\dagger_1\eta_1+\tilde\eta^\dagger_1\tilde\eta_1)\Big]
\eea
to the interaction terms
\be
  \frac{2\pi i}{k}\frac{N(N-1)}{2}\Big[(|\phi_1|^2+|\chi_1|^2)(\eta^\dagger_2\eta_2+\tilde\eta^  
  \dagger_2\tilde\eta_2)
  +(|\phi_2|^2+|\chi_2|^2)(\eta^\dagger_1\eta_1+\tilde\eta^\dagger_1\tilde\eta_1)\Big]\,.
\ee
The fermion kinetic terms, on the other hand, contribute
\be
-i\Tr\Big[\psi^{\dagger A}(\gamma^\mu D_\mu +\mu)\psi_A\Big]=-i\frac{N(N-1)}{2}\sum_{i=1,2}\Big[\eta_i^\dagger (\Dslash +\mu)\eta_i +
\tilde\eta_i^\dagger(\Dslash+\mu)\tilde\eta_i\Big]\,,
\ee
so that the total effective {\it abelian} action is then
\bea
S&=&-\frac{N(N-1)}{2}\int d^3x\Bigg\{\frac{k}{4\pi}\epsilon^{\mu \nu \lambda}\big(a^{(2)}_{\mu}f^{(1)}_{\nu   
  \lambda}+a^{(1)}_{\mu}f^{(2)}_{\nu \lambda}\big)
  +|D_{\mu}\phi_{i}|^{2}+|D_{\mu}\chi_{i}|^{2}\cr
&&+\sum_{i=1,2}\Big[\eta_i^\dagger (\Dslash +\mu)\eta_i +
\tilde\eta_i^\dagger(\Dslash+\mu)\tilde\eta_i\Big]\cr
&&-\frac{2\pi i}{k}\Big[(|\phi_1|^2+|\chi_1|^2)(\eta^\dagger_2\eta_2+\tilde\eta^\dagger_2\tilde\eta_2)
+(|\phi_2|^2+|\chi_2|^2)(\eta^\dagger_1\eta_1+\tilde\eta^\dagger_1\tilde\eta_1)\Big]\cr
&&+\left(\frac{2\pi}{k}\right)^2\Big[(|\phi_{1}|^{2}+|\chi_{1}|^{2})\big(|\chi_{2}|^{2}-|\phi_{2}|^{2}
  -c^{2}\big)^{2}
  +(|\phi_{2}|^{2}+|\chi_{2}|^{2})\big(|\chi_{1}|^{2}-|\phi_{1}|^{2}-c^{2}\big)^{2}\cr
&+&4|\phi_{1}|^{2}|\phi_{2}|^{2}(|\chi_{1}|^{2}+|\chi_{2}|^{2})+4|\chi_{1}|^{2}|\chi_{2}|^{2}(|\phi_{1}|^{2}+|\phi_{2}|^{2})\Big]\Bigg\}
\label{susyaction}
\eea
Having started off with an $\mathcal{N}=6$ supersymmetric theory in $(2+1)-$dimensions, at this point it is worth asking how much (if any) of the supersymmetry is preserved by our abelianization procedure.
  
\subsection{Supersymmetry}

The susy rules for the massless ABJM model are \cite{Gomis:2008vc,Chu:2010fk}
\bea
\delta C^A&=& i\bar \epsilon^{AB}\psi_B,\nonumber\\
\delta \psi_B&=&\gamma^\mu \epsilon_{AB}D_\mu C^A+\frac{2\pi}{k}2C^CC_B^\dagger C^D \epsilon_{CD}-\frac{2\pi}{k}(C^AC^\dagger_CC^C-C^CC^\dagger_CC^A)\epsilon
_{AB},\nonumber\\
\delta A_\mu &=&-\frac{2\pi}{k}(\bar \epsilon_{AB}\gamma_\mu C^B\psi^{\dagger A}-\bar \epsilon^{AB}\gamma_\mu \psi_A C^\dagger_B),\\
\delta\hat A_\mu&=&-\frac{2\pi}{k}(\bar \epsilon_{AB}\gamma_\mu \psi^{\dagger A}C^B-\bar \epsilon^{AB}\gamma_\mu C_B^\dagger \psi_A),\nonumber
\eea
where indices are raised and lowered with the $SU(4)$ invariant metric $\delta_A^B$, and the susy parameters $\epsilon^{AB}$ are antisymmetric in $AB$ and 
satisfy
\be
\epsilon^{AB}=\frac{1}{2}\epsilon^{ABCD}\epsilon_{CD},
\ee
which means they live in the ${\bf 6}$ representation of $SU(4)$.

The mass deformation introduces an extra term in the fermion supersymmetry transformation rules \cite{Bagger:2012jb}
\be
\delta^{(\mu)} \psi_A=\frac{1}{2}\epsilon_{DF}C^F{\begin{pmatrix}\mu &0&0&0\\0&\mu&0&0\\0&0&-\mu&0\\0&0&0&-\mu\end{pmatrix}_A}^D\,,
\ee
where the normalization is such that the scalar mass term in the Lagrangian is $-\mu^2\Tr[\bar C_A C^A]$.

The reality condition for the supersymmetry parameter
means (in $\a,\dot\a$ components) that the independent components of $\epsilon^{AB}$ are $(\epsilon^{12},\epsilon^{1\dot 1}, 
\epsilon^{1\dot 2})$, which are complex, i.e.  ${\cal N}=6$ real supersymmetries. The other components are related by the reality condition
to these ones as:
\be
\epsilon^{\dot 1\dot 2}=\epsilon^{12};\;\;\;
\epsilon^{2\dot 2}=-\epsilon^{1\dot 1};\;\;\;
\epsilon^{\dot 1 2}=-\epsilon^{1\dot 2},
\ee
and as before, the $\epsilon$'s are antisymmetric, so for example, $\epsilon^{\dot 1 1}=-\epsilon^{1\dot 1}$, etc. 

We now split the susy laws in $\a,\dot\a$ components as well, and substitute the full abelianization ansatz,
\bea
Q^\a&=&\phi_\a G^\a,\nonumber\\
R^{\dot\a}&=&\chi_{\dot\a}G^{\dot\a},\nonumber\\
A_\mu&=&a_\mu^{(2)}G^1G^\dagger_1+a_\mu^{(1)} G^2 G^\dagger_2, \nonumber\\
\hat A_\mu&=&a_\mu^{(2)}G^\dagger_1 G^1+a_\mu^{(1)}G^\dagger_2 G_2,\\
\psi_\a&=&\eta_\a G^\a,\nonumber\\
\tilde\psi_{\dot\a}&=&\tilde \eta_{\dot\a}G^{\dot\a},\nonumber
\eea

To understand what will happen, we write out explicitly one component, namely $Q^1$, for which we obtain
\be
\left(\delta \phi_1\right)G^1=i\bar \epsilon^{12}\left(\eta_2 G^2\right)+i\bar\epsilon^{1\dot 1}
\left(\tilde \eta_{\dot 1} G^{\dot 1}\right)+i\bar\epsilon^{1\dot 2}\left(\tilde \eta_{\dot 2}
G^{\dot 2}\right),
\ee
For this to make sense, {\it i.e.} for this susy transformation law to remain a symmetry after the abelianization, the right hand side needs to be proportional to $G^1$ also. This restricts us to 
$\epsilon^{12}=\epsilon^{1\dot 2}=0$, and only $\epsilon^{1\dot 1}\neq 0$. Repeating the argument for the other scalar components, we find that 
\bea
(\delta \phi_1)G^1&=& (i\bar\epsilon^{1\dot 1}\tilde\eta_{\dot 1})G^{\dot 1},\nonumber\\
(\delta \phi_2)G^2&=& (i\bar\epsilon^{2\dot 2}\tilde\eta_{\dot 2})G^{\dot 2},\\
(\delta \chi_{\dot 1})G^{\dot 1}&=& (i\bar\epsilon^{\dot 1 1}\eta_{1})G^{1},\nonumber\\
(\delta \chi_{\dot 2})G^{\dot 2}&=& (i\bar\epsilon^{\dot 2 2}\eta_{2})G^{2},\nonumber
\eea
or, using the relations between the epsilon components and peeling off the G matrices, 
\bea
\delta \phi_1&=& i\bar\epsilon^{1\dot 1}\tilde\eta_{\dot 1},\nonumber\\
\delta \phi_2&=& -i\bar\epsilon^{1\dot 1}\tilde\eta_{\dot 2},\\
\delta \chi_{\dot 1}&=& -i\bar\epsilon^{1\dot 1}\eta_{1},\nonumber\\
\delta \chi_{\dot 2}&=& i\bar\epsilon^{1 \dot 1}\eta_{2}.\nonumber
\eea
Clearly then, for the scalars only the $\epsilon^{1\dot 1}$ independent component is nonzero.

For the gauge fields, substituting the abelianization ansatz in the transformation law 
for $A_\mu$, gives
\bea
\left(\delta a_\mu^{(2)}\right)G^1G^\dagger_1+\left(\delta a_\mu^{(1)}\right)G^2G^\dagger_2&=&-\frac{2\pi}{k}\Bigg[\bar \epsilon_{1\dot 1}\gamma_\mu
\left(Q^1\tilde \psi^{\dagger \dot1 }-R^{\dot 1}\psi^{\dagger 1}\right)+\bar \epsilon_{2\dot 2}\gamma_\mu\left(Q^2\tilde
\psi^{\dagger \dot 2}-R^{\dot 2}\psi^{\dagger 2}\right)\cr
&-&\bar \epsilon^{1\dot 1}\gamma_\mu\left(\psi_1R^\dagger_{\dot 1}-\tilde \psi_{\dot 1}Q^\dagger _1\right)-\bar \epsilon^{2\dot 2}\gamma_\mu \left(\psi_2R^\dagger_{\dot 2}-\tilde \psi_{\dot 2}Q^\dagger_2\right)\Bigg],
\eea
where we have already kept only the terms proportional to $G^1G^\dagger_1$ or $G^2G^\dagger_2$ on the right hand side, and set to zero the rest. Again only the independent $\epsilon^{1\dot 1}$ component survives. Identifying the coefficients of 
$G^1G^\dagger_1$ and $G^2G^\dagger_2$ on the left hand side and the right hand side, we find that
\bea
  \delta a_\mu^{(1)}&=&\frac{2\pi}{k}\bar \epsilon_{1\dot 1}\gamma_\mu[\phi_2\tilde\eta_{\dot 2}  
  ^*+\phi_2^*\tilde \eta_{\dot 2}-\chi_{\dot 2}\eta_2^*-\chi^*_{\dot 2}
  \eta_2],\nonumber\\
  &&\\
  \delta a_\mu^{(2)}&=&-\frac{2\pi}{k}\bar \epsilon_{1\dot 1}\gamma_\mu[\phi_1\tilde\eta_{\dot   
  1}^*+\phi_1^*\tilde \eta_{\dot 1}-\chi_{\dot 1}\eta_1^*
  -\chi^*_{\dot 1}\eta_1].\nonumber
\eea
Next, we move on to the fermion rules. Here again we keep only the independent $\epsilon^{1\dot 1}$ component (having checked that the rest give a different 
matrix dependence on the right hand side from the left hand side), and find for $\psi_{1}$, that
\be
(\delta \eta_1)G^1=\gamma^\mu \epsilon_{1\dot 1}D_\mu R^{\dot 1}-\frac{2\pi}{k}\Big[R^{\dot 1}(Q_2^\dagger Q^2+R^\dagger_{\dot 2}R^{\dot 2})-
(Q_2^\dagger Q^2+R^\dagger_{\dot 2}R^{\dot 2})R^{\dot 1}\Big]\epsilon_{\dot 1 1}+\frac{1}{2}\mu \epsilon_{1\dot 1}R^{\dot 1}.
\ee
Using the relations
\be
G^1=G^1 G^\dagger_2 G^2-G^2 G^\dagger_2 G^1;\;\;\;\;
G^2=G^2G_1^\dagger G^1-G^1 G^\dagger_1 G^2,
\ee
peeling off $G^1$ and using the relations between epsilons, we get
\be
\delta \eta_1=\gamma^\mu \epsilon_{1\dot 1}D_\mu \chi_{\dot 1}+\frac{2\pi}{k}\epsilon_{1\dot 1}\chi_{\dot 1}(|\phi_2|^2+|\chi_{\dot 2}|^2)+\frac{\mu}{2}
\epsilon_{1\dot 1}\chi_{\dot 1}
\ee
Repeating for $\psi_2,\psi_{\dot 1},\psi_{\dot 2}$, we get
\bea
\delta \eta_2&=&-\gamma^\mu \epsilon_{1\dot 1}D_\mu \chi_{\dot 2}-\frac{2\pi}{k}\epsilon_{1\dot 1}\chi_{\dot 2}(|\phi_1|^2+|\chi_{\dot 2}|^2)-\frac{\mu}{2}
\epsilon_{1\dot 1}\chi_{\dot 2},\nonumber\\
\delta \tilde \eta_{\dot 1}&=&-\gamma^\mu \epsilon_{1\dot 1}D_\mu \phi_{ 1}-\frac{2\pi}{k}\epsilon_{1\dot 1}\phi_{1}(|\phi_2|^2+|\chi_{\dot 2}|^2)+\frac{\mu}{2}
\epsilon_{1\dot 1}\phi_{ 1},\\
\delta \tilde \eta_{\dot 2}&=&\gamma^\mu \epsilon_{1\dot 1}D_\mu \phi_{ 2}+\frac{2\pi}{k}\epsilon_{1\dot 1}\phi_{2}(|\phi_1|^2+|\chi_{\dot 1}|^2)-\frac{\mu}{2}
\epsilon_{1\dot 1}\phi_{2}.\nonumber
\eea
Finally then, for ease of notation, renaming $\epsilon^{1\dot 1}$ as simply $\epsilon$, we write down the susy transformation rules
\bea
\delta \phi_1&=& i\bar\epsilon\tilde\eta_{\dot 1},\nonumber\\
\delta \phi_2&=& -i\bar\epsilon\tilde\eta_{\dot 2},\nonumber\\
\delta \chi_{\dot 1}&=& -i\bar\epsilon\eta_{1},\nonumber\\
\delta \chi_{\dot 2}&=& i\bar\epsilon\eta_{2},\nonumber\\
\delta a_\mu^{(1)}&=&\frac{2\pi}{k}\bar \epsilon\gamma_\mu[\phi_2\tilde\eta_{\dot 2}^*+\phi_2^*\tilde \eta_{\dot 2}-\chi_{\dot 2}\eta_2^*-\chi^*_{\dot 2}
\eta_2],\nonumber\\
\delta a_\mu^{(2)}&=&-\frac{2\pi}{k}\bar \epsilon\gamma_\mu[\phi_1\tilde\eta_{\dot 1}^*+\phi_1^*\tilde \eta_{\dot 1}-\chi_{\dot 1}\eta_1^*
-\chi^*_{\dot 1}\eta_1],\\
\delta \eta_1&=&\gamma^\mu D_\mu \chi_{\dot 1}+\frac{2\pi}{k}\epsilon\chi_{\dot 1}(|\phi_2|^2+|\chi_{\dot 2}|^2)+\frac{\mu}{2}\epsilon\chi_{\dot 1},\nonumber\\
\delta \eta_2&=&-\gamma^\mu \epsilon D_\mu \chi_{\dot 2}-\frac{2\pi}{k}\epsilon\chi_{\dot 2}(|\phi_1|^2+|\chi_{\dot 2}|^2)-\frac{\mu}{2}\epsilon\chi_{\dot 2},\nonumber\\
\delta \tilde \eta_{\dot 1}&=&-\gamma^\mu \epsilon D_\mu \phi_{ 1}-\frac{2\pi}{k}\epsilon\phi_{1}(|\phi_2|^2+|\chi_{\dot 2}|^2)+\frac{\mu}{2}\epsilon\phi_{ 1},\nonumber\\
\delta \tilde \eta_{\dot 2}&=&\gamma^\mu \epsilon D_\mu \phi_{ 2}+\frac{2\pi}{k}\epsilon\phi_{2}(|\phi_1|^2+|\chi_{\dot 1}|^2)-\frac{\mu}{2}\epsilon\phi_{2}.\nonumber
\label{susyrules}
\eea
Since the parameter $\epsilon$ is complex, we have an $SO(2)=U(1)$ R-symmetry and this is therefore ${\cal N}=2$ susy in 3 dimensions. 

An immediate question that arises is whether we can supersymmetrize the further truncation to the Landau-Ginzburg system in \cite{Mohammed:2012rd,Mohammed:2012gi}. 
From the above rules, we see that setting $\phi_1=\phi_2=0$ and $\chi_{\dot 1}=b$ 
requires that $\tilde \eta_{\dot 1}=\tilde \eta_{\dot 2}=0$ for consistency. However, because of the relation $\delta \chi_{\dot 1}=-i\bar\epsilon \eta_1$ this would seem to also require $\eta_1=0$. With this in place, most things are consistent, except the relation for $\delta \eta_1$, whose right hand side gives the consistency condition
\be
b\Big[+i\gamma^\mu a_\mu^{(1)}\epsilon-\frac{2\pi}{k}|\chi_{\dot 2}|^2\epsilon +\frac{\mu}{2}\epsilon\Big]=0
\ee
which is a constraint on fields and so is not satisfied in general. Apparently then, the Landau-Ginzburg system can {\it not} be supersymmetrized in general.

\subsection{Global charge analysis and applicability to condensed matter}

To summarise our findings of the previous section, the reduced effective action in (\ref{susyaction}) has ${\cal N}=2$ supersymmetry. It has a {\it local} $U(1)\times U(1)$ invariance, where the fields $\phi_1,\chi_1,\eta_1,\tilde\eta_1$ are charged with charge $+1$ under the first $U(1)$, and $\phi_2,\chi_2,\eta_2,\tilde\eta_2$ all carry charge $+1$
under the second $U(1)$.

With respect to {\it global} symmetries, the original ABJM model had $SU(4)\times U(1)$ R-symmetry before the addition of the mass term. Adding the mass term breaks this 
to $SU(2)\times SU(2)\times U(1)_A\times U(1)_B\times {\mathbb Z}_2$, where the $SU(2)$'s act on $Q$ and $R$ respectively. Specifically, $U(1)_A$ acts on $Q$ with charge
$+1$ and on $R$ with charge $-1$, and ${\mathbb Z}_2$ interchanges $Q$ and $R$. 

We note now that the action (\ref{susyaction}) has an overall $U(1)^8\times {\mathbb Z}_2$ global invariance. Each of the $U(1)$'s acts on just one of the eight fields 
$\phi_1,\phi_2,\chi_1,\chi_2,\eta_1,\eta_2,\tilde\eta_1,\tilde\eta_2$ and not on any of the rest. Out of these global symmetries, two linear 
combinations defined above are promoted to local invariances, leaving a total of six factors of $U(1)$ as just global invariances. In addition, the ${\mathbb Z}_2$ 
acts by interchanging indices 1 and 2, i.e.
\be
\phi_1\leftrightarrow \phi_2,\, \chi_1\leftrightarrow \chi_2,\,\eta_1\leftrightarrow \eta_2,\,\tilde\eta_1\leftrightarrow \tilde\eta_2,\, a_\mu^{(1)}\leftrightarrow
a_\mu^{(2)}.
\ee
Compared to the action in (\ref{hsmodel}), here the gauge charge is either $+1$ or zero for both groups (for half of the fields it is $+1$ and the other 
half, 0), which is already different. We could, of course, choose the equivalent of the global electric charge of (\ref{hsmodel}) to be, completely diagonal,
\be
Q\sim \phi_i^\dagger \phi_i+\chi_i^\dagger\chi_i+\eta_i^\dagger \eta_i+\tilde \eta_i^\dagger \tilde \eta_i
\ee
but this would be just the sum of the two gauge charges (for the two $U(1)$ gauge groups), and so would not constitute an independent charge. We could also have considered that
the field $\phi$, for example, carries charge $+1$ and $\phi^\dagger$, $-1$, but that would only mean that there are positive and negative charges, and we would still have 
a global charge that is the sum of the two local charges.

As an alternative, one could choose as the electric charge some other combination of the eight global $U(1)$'s like, for example, the charge under which the $\phi_i,\eta_i$ have 
charge $+1$, and $\chi_i,\tilde\eta_i$ have charge $-1$. Then, considering that the local charges are $+1$ for the fields and $-1$ for their conjugates, 
the global charge is different from any linear combination of the local charges. However this global charge has both positive and negative carriers, unlike the expression in (\ref{hselcharge}).

In conclusion, while our abelianization produces a model that exhibits some of the more general features of the condensed matter model (\ref{hsmodel}), it differs in some of the finer details and since, as they say, this is where the devil is, it is worth exploring other avenues as well.

\section{Adding fundamentals to the ABJM model}

To this end, we now consider the addition of fundamental fields to the ABJM model, with the ultimate goal of specializing to $N=1$ at the end, and thus obtaining the required charged fields even in the abelian case. In the case of $N_c$ coincident $Dp-$branes, with an $SU(N_c)$ gauge theory living on them, adding fundamentals can be done in two ways. The first is via the addition of a probe $D(p+4)-$brane (or $D(p+2)-$brane) to the gravitational background set up by the $N_c$ $Dp-$branes with the role of the fundamental scalars in the $SU(N_{c})$ gauge theory played by open strings stretched between the stack of $Dp-$branes and the probe brane. In this case however, a $Dp-D(p+4)$ system with the $D(p+4)-$brane wrapped on a compact space is not consistent, since the flux has nowhere to go on a compact 
space\footnote{As a simple intuitive picture, a positive electric charge on a circle has flux lines going away from the charge on both sides, so it needs a corresponding 
negative charge somewhere else to sink the flux lines incoming from both sides.}. To cancel the flux, one needs to add negative charge. Normally, this would require the addition of anti-branes, but if one wants to preserve supersymmetry this is no longer an option. In this case, the only consistent solution is to add an orientifold plane, for a total system of $Dp-D(p+4)-O(p+4)$.

A further complication in our case is that the ABJM model has a {\it product} gauge group, $SU(N)\times SU(N)$, with {\it bifundamental} scalars and fermions under both 
gauge groups, so it is not obvious at all how to add the fundamentals. Fortunately, all is not lost since we do know that giving a VEV to one of the scalars, say $\langle X^1\rangle=v$, turns the 
ABJM model into a $D2-$brane gauge theory with $SU(N)$ gauge group, via a 3d Higgs mechanism \cite{Mukhi:2008ux,Chu:2010fk}. Thereafter, the construction should reduce to the general one.

\subsection{A Review of the D3-D7-O7 case}

To better understand how to proceed, let's begin by reviewing the case of $D3-$branes. In this case, consistently adding fundamental fields while 
preserving supersymmetry is achieved by 
adding $D7-$branes and an orientifold $O7-$plane, as in the construction of \cite{Aharony:1998xz}. The $D3-$branes by themselves, of course, provide an $SU(N)$ gauge group, and the gravity dual is the all too familiar $AdS_5\times S^5$. Adding $D7-$branes to this background
means that they need to be parallel to $AdS_5$ and wrapping a 3-cycle inside the $S^5$. That means that for consistency of the 7-brane flux, we need to add an orientifold plane, which acts as a sink for the flux. This orientifolding means that the gauge group is now $USp(2N)$ and we have an ${\cal N}=2$ superconformal field theory. Since the orientifold plane carries charge -4, one needs to add four $D7-$branes together with the $O7-$plane, resulting in a global $SO(2N_f)=SO(8)$ group.

\subsubsection{The Gravity Picture}
For the gravity dual, the spatial $\mathbb{Z}_2$ orbifold part of the orientifold acts on $AdS_5\times S^5$ as follows. The metric on $AdS_5\times S^5$ is
written as
\bea
&&ds^2=R^2(-\cosh^2\rho dt^2+d\rho^2+\sinh^2\rho d\Omega_3^2+\cos^2\theta d\psi^2+d\psi^2
+\sin^2\theta d\tilde{\Omega}_3^2)\nonumber\\
&&d\tilde{\Omega}_3^2=\cos^2\theta 'd{\psi '}^2+d{\theta '}^2+\sin^2\theta 'd\phi ^2,\label{o7metric}
\eea
and then the $\mathbb{Z}_2$ acts by sending $\psi '\rightarrow \psi '+\pi$. The effect of the orientifolding is to have $\theta '\in (0,\pi/2)$ instead 
of $(0,\pi)$, and the invariant plane (the orientifold $O(7)$ plane) is situated at $\theta '=\pi/2$ and carries 
$-4$ units of $D7$-brane charge. It is this that is cancelled by the addition of the four $D7$-branes. At this point, it is worth noticing two things: First, the plane is actually an $S^3$ inside $S^5$ and second, that we could also have chosen the plane at $\theta=\pi/2$ instead of the one at $\theta'=\pi/2$ as the $O7-$plane.

Since the decoupling limit leaves the $D7-$branes unchanged, there are the four $D7-$branes on top of the $O7-$plane wrapping the $S^3$. This results in a $(7+1)-$dimensional Super Yang-Mills theory at the location of the $O7-$plane in the gravity dual. The KK modes of the 8- dimensional SYM reduced on $S^3$ 
are fields in a representation of the $SO(4)$ group, and charged under the global $SO(2N_f)=SO(8)$ group. 

\subsubsection{The Field Theory Picture}
In order to establish the line of argument as well as our notation, let's briefly review the analysis presented in \cite{Aharony:1998xz,Berenstein:2002zw}. The ${\cal N}=4$ SYM multiplet on $N$ $D3-$branes is composed of $(A_\mu ^a, \psi_\a^{aI},Z^{a[IJ]})$, where $a$ is an index in the adjoint of $U(N)$, which can also be written as $a=i\bar j$, with 
$i,\bar j=1,...,N$ in the (anti)fundamental of $U(N)$.
The $Z$ are complex scalars, with $I=1,...,4$ an index in the fundamental of $SU(4)$ (or equivalently the spinor of $SO(6)$), meaning that
$[IJ]$ is in the antisymmetric {\bf 6} representation of $SU(4)$. These fields satisfy the reality condition 
\be
Z^{a[IJ]}=\epsilon^{IJKL}(Z^\dagger)^a_{KL},
\ee
and correspond to the six real transverse coordinates of the $D3-$brane.

Orientifolding corresponds to identifying some of the fields. Under this identification, the gauge group changes from $SU(N)$ to $USp(2N)$. Importantly, adjoint fields remain in the 
adjoint,  except that instead of an unrestricted $ij$, the adjoint of $USp(2N)$ is the symmetric $(ij)$, or $2N(2N+1)/2$ representation. The scalars 
in the adjoint representation now are simply singlets under the remaining $SU(2)\times SU(2)$ global symmetry, and carry unit charge under the $U(1)$.

Adding the $D7-$branes means that we also need an antisymmetric field $Y_{[ij]}^{AA'}$ in the $2N(2N-1)/2$ representation as well as four fundamentals, $q_i^{Am}$, 
of the gauge group $USp(2N)$. This breaks supersymmetry down to ${\cal N}=2$, which in turn means that the R-symmetry splits as 
$SO(6)\rightarrow SO(4)\times SO(2)$, or equivalently $SU(4)\rightarrow SU(2)\times SU(2)\times U(1)$. Of this, only the $SU(2)_R\times U(1)$ is still an R-symmetry and 
the other, $SU(2)_L$, factor becomes a simple global symmetry. 
The index $I$ splits into $AA'$, with $A,A'=1,2$
corresponding to the two $SU(2)$ factors. While $Y^{AA'}_{[ij]}$ respects the $SU(4)$ symmetry,  $q_i^{Am}$ doesn't, since it is charged only under 
$SU(2)_R$, which is the only R-symmetry of the final model. The index $m=1,...,8$ belongs to the flavor group $SO(8)$. 

In summary, 
\begin{itemize}
  \item $Z$ is a singlet of $SU(2)_L,SU(2)_R$ and $SO(8)$, with unit charge under $U(1)$; 
  \item $Y$ is a doublet of $SU(2)_L$ and a doublet of $SU(2)_R$ and 
  singlet under $U(1)$ and 
  \item $q$ is a singlet under $SU(2)_L$, a doublet under $SU(2)_R$, 
  has no charge under $U(1)$ and is in the fundamental of $SO(8)$.
\end{itemize}
The fields $Y$ and $q$ satisfy reality conditions
\bea
&&q_i^A=\epsilon^{AB}J_{ij}(q^\dagger)^j_B\nonumber\\
&&\\
&&Y_{[ij]}^{AA'}=\epsilon^{AB}\epsilon^{A'B'}J_{ik}J_{jl}(Y^\dagger)^{[kl]}_{BB'}\nonumber
\eea
where $J_{ij}$ is the antisymmetric invariant matrix of $USp(2N)$,
\be
J_{ij}=\begin{pmatrix} 0&{\mathbf 1}\\-{\mathbf 1}&0\end{pmatrix}
\ee
Because of the reality conditions, $q^A$ are two real scalars, corresponding to the two {\it overall} transverse coordinates, {\it i.e.} transverse to both the $D3-$branes
and the $D7-$branes (as do the two real components of the complex adjoint $Z$ scalars), 
and $Y^{AA'}$ are four real scalars, corresponding the four {\it relative} transverse coordinates, namely parallel to the $D7-$branes, but transverse to the $D3-$branes.

In ${\cal N}=1$ language, where we can write the scalars $q_i^{Am}$ as belonging to 
two chiral superfields $q^m$ and $\tilde q^m$, and the scalars $Y^{AA'}_{[ij]}$ as belonging to two chiral superfields $Y^{A'}$ and $Y'^{A'}$, the superpotential can be written as 
\be
{\cal W}\sim (Z_{ij}q^{im}\tilde q^{jm}+Z_{ij}J^{jk}Y^{A'}_{kk'}J^{k'l}Y'^{A'}_{ll'}J^{l'i}),
\ee
with the resulting F-terms 
\bea
F_{\tilde q}&=&Zq,\nonumber\\
F_q&=&Z\tilde q,\nonumber\\
F_{Y'\;\; il}&=&Z_{ij}J^{jk}Y_{kl}-(i\leftrightarrow l),\\
F_{Y\;\; il}&=&Z_{ij}J^{jk}Y'_{kl}-(i\leftrightarrow l),\nonumber\\
F_{Wil}&=&Y_{ij}J^{jk}Y'_{kl}+(i\leftrightarrow l)+(q\tilde q )_{(il)}.\nonumber
\eea
For the component action, the scalar potential is given by the squares of the F-terms plus the square of the D-terms.

To close this discussion of the field theory side of the orientifolding, we point out that 
\begin{itemize}
  \item there are operators with $SO(4)_R=SU(2)\times SU(2)$ indices as well as flavor 
  $SO(8)$ indices, like for instance ${\cal O}^{ABmn}=\bar q^{Am}q^{Bn}$ 
  (with an implicit sum over the gauge indices $i$), which couple to the fields coming from 
  the $S^3$ reduction of 7+1 dimensional SYM in the gravity dual and 
  \item there are also the usual gauge invariant operators without flavor indices, only 
  with $SO(6)_R=SU(4)_R$ indices, which couple to the fields coming from the $S^5$ 
  reduction of supergravity fields in the gravity dual.
\end{itemize}

\subsection{The Gravity Dual}
Returning to our case at hand,  we consider a construction which should reduce to a D2-D6-O6 system when we reduce the ABJM model down to type IIA string theory, i.e. a 
T-dual construction to the one above. The background that we work with is $AdS_4\times \mathbb{CP}^3$ instead of $AdS_5\times S^5$ but, as before, 
we look for a $\mathbb{Z}_2$ symmetry leading to an $O6-$plane inside the gravity dual. 

The Fubini-Study metric on ${\mathbb CP}^3$ is 
\bea
ds^2&=&d\xi^2+\frac{\cos^2\xi}{4}(d\theta_1^2+\sin^2\theta_1d\phi_1^2)+\frac{\sin^2\xi}{4}(d\theta_2^2+\sin^2\theta_2d\phi_2^2)\nonumber\\
&&+\cos^2\xi\sin^2\xi(d\psi+A_1-A_2),\\
A_i&=&\frac{1}{2}\cos\theta_id\phi_i.\nonumber
\eea
Just as in the $D3-D7$ case, we will choose the $O6-$plane parallel to $AdS_4$ and wrapping the codimension-3 cycle in $\mathbb{CP}^3$,
\be
\theta_1=\theta_2=\frac{\pi}{2};\;\;\;\;
\psi=\pi,
\ee
which is a fixed plane for the $\mathbb{Z}_2$ action
\be
\phi_1\rightarrow \phi_1+\pi;\;\;\;
\phi_2\rightarrow \phi_2+\pi;\;\;\;
\psi\rightarrow \psi+\pi\label{z2}.
\ee
The metric on this fixed plane is  
\be
ds^2=d\xi^2+\frac{\cos^2\xi}{4}d\phi_1^2+\frac{\sin^2\xi}{4}d\phi_2^2,
\ee
with coordinate ranges $0<\xi<\pi/2$ and $0\leq \phi_i\leq 2\pi$ and the action (\ref{z2}), which is an $S^3/\mathbb{Z}_2\in \mathbb{CP}^3$. The same cycle has been used in \cite{Hikida:2009tp}, where more details of the construction are given, so we will wait until we compare to that paper to comment further. We note also that in \cite{Hikida:2009tp}, it was observed that this construction makes an orbifold, but it was not required that an orientifold
$O6-$plane live there, as we do.

\subsection{The Field Theory}

We start by trying to understand the system when thought of as a D2-D6-O6 system, {\it i.e.} when one of the scalars has aquires VEV and the Higgs mechanism on 
the ABJM model leads to $N$ D2-branes. That analysis is T-dual to the D3-D7-O7 case studied above, so it will be quite similar.

\subsubsection{D2-brane Analysis}

The field theory living on the stack of $N$ $D2-$branes is an ${\cal N}=8$ Super Yang-Mills, with field content $(A_\mu^a,\psi_\a^{aA},X^{aI'})$, where
$a=ij$ is in the adjoint of $SU(N)$, $I'=1,...7$ is in the fundamental of $SO(7)$ 
and $A=1,...,4$ is in the fundamental of $SU(4)$ or spinor of $SO(6)$, which can be also understood as the spinor of $SO(7)$. We have an overall $SO(8)$ R-symmetry, of which only $SO(7)_R$ is manifest. 

When we do the orientifold projection, the gauge group changes from $SU(N)$ to $USp(2N)$, and the fields are still in the adjoint, though the adjoint of 
$USp(2N)$ is now a symmetric representation, $a=(ij)\in 2N(2N+1)/2$. However, the scalar fields $X$ are now restricted to be in a three 
dimensional (adjoint) representation of the $SU(2)\times SU(2)$ relative transverse global symmetry.

The addition of the $D6-$branes again breaks $SO(7)$ to $SO(4)\times SO(3)$, or $SU(2)\times SU(2)\times SU(2)$, and splits the index $A$ into 
$MM'$ under the two $SU(2)$ factors. We will also have an $SO(8)$ flavor group, with fundamental index $m=1,...,8$, because, again, we need to add four $D6-$branes
to cancel the -4 charge of the $O6-$plane. As before, we also need to add antisymmetric tensor fields $Z^{MM'}_{[ij]}$ in the $2N(2N-1)/2$ representation
of $USp(2N)$, which satisfy the same reality condition as before, 
\be
Z_{[ij]}^{MM'}=\epsilon^{MN}\epsilon^{M'N'}J_{ik}J_{jl}(Z^\dagger)^{[kl]}_{NN'},
\ee
meaning that again we have four scalars corresponding to the four relative transverse directions, parallel to the $D6-$branes and transverse to the $D2-$branes.

Finally, we need to add fundamental scalars $q_i^{MM''m}$ coming from the strings stretching between the $D2-$branes and the $D6-$branes, and corresponding to four 
overall transverse coordinates (transverse to both $D2-$branes and $D6-$branes). Note that naively there should be {\it three} coordinates, but because of supersymmetry, the chiral multiplet has to have {\it four} scalars. This fits quite nicely with our interpretion of the following model as coming from eleven dimensions where it will be clear that we need four scalars. 
So, unlike the $D3-D7$ case now the number of $X$'s (3) differs from the number of $q$'s (4), pointing to the fact that there is a better interpretation in eleven dimensions.

Here the index $M$ is in a two-dimensional representation of one of the relative transverse $SU(2)$'s, while $M''$ is in a two-dimensional 
representation of the overall transverse $SU(2)$. Since the scalars need to be four, they satisfy a reality condition,
\be
  q_i^{MM''m}=J_{ij}\epsilon^{MN}\epsilon^{M''N''}(q^\dagger)^{mj}_{NN''}.
\ee
We will write the superpotential in the ABJM case only, since it is easier to understand, and this case can be obtained by Higgsing.

\subsubsection{Lifting to the ABJM Model}

The ABJM model is an ${\cal N}=6$ supersymmetric model with gauge group $U(N)\times U(N)$ and gauge fields $A_\mu^{a(1)},A_\mu^{a'(2)}$ and bifundamental 
fields $Z^{Aii'}$ and $\psi_\a^{Aii'}$, where $a$ and $a'$ are in the adjoint of $U(N)$, while
$i,i'=1,...,N$ are in the fundamental of $U(N)$, and $A=1,...,4$ is in 
the fundamental of $SU(4)_R$. The R-symmetry group is $SU(4)\times U(1)=SO(6)\times SO(2)$.

The orientifold projection changes the adjoints $a,a'$ into symmetric tensor adjoints $a=(ij)$ and $a'=(i'j')$ of $USp(2N)$. However, it now also requires that we identify the two gauge group factors. 

To see how this works, it will be useful to review how the ABJM model is constructed. We start with $N$ $D2-$branes in type IIA wrapping a compact direction, and broken in two places by an $NS5-$brane and a $NS5'-$brane. Then one adds $k$ $D6$-branes
to one of the $NS5-$branes to turn it into a $(1,k)$ 5-brane and rotates. The rest of the procedure is not important for our discussion. Bifundamental fields arise from 
strings stretching between one half of the $D2-$branes to the other half, through the $5-$brane. Orientifolding corresponds to adding an $O6-$plane at a certain 
point in the compact direction. This can only be the location of one of the $5-$branes because of symmetry (since otherwise we would get a set-up which is 
asymmetric between the two gauge groups). That means that the orientifold projection will identify the two half-branes, i.e. the two gauge groups. 

The identification means that now we can decompose $(ii')$ in irreducible representations of the unique gauge group, i.e.
\be
Z^{Aii'}=Z^{A(ii')}+Z^{A[ii']}
\ee
where the symmetric tensor is the adjoint of $USp(2N)$, and $[ii']$ is the antisymmetric representation in the decomposition. The orientifold projection will also impose the reality conditions
\bea
Z^{MM'[ii']}&=&\epsilon^{MN}\epsilon^{M'N'}J_{ij}J_{i'j'}(Z^\dagger)^{[jj']}_{NN'},\nonumber\\
\\
Z^{MM'(ii')}&=&\epsilon^{MN}\epsilon^{M'N'}J_{ij}J_{i'j'}(Z^\dagger)^{(jj')}_{NN'},\nonumber
\eea
which means that the $Z^{A[ii']}$ now describe the four scalars corresponding to the relative transverse directions (parallel to the $D6-$branes, but transverse to 
the $D2-$branes), whereas $Z^{A(ii')}$ describe as usual the four scalars corresponding to the overall transverse directions (transverse to both $D2-$branes and 
$D6-$branes).

Adding $D6-$branes breaks the R-symmetry from $SU(4)\times U(1)$ to $SU(2)\times SU(2)\times U(1)$, while adding fundamental fields coming from the strings stretching between the $D6-$branes and the two halves of the $D2-$branes. Naively, this leads 
to fields corresponding to four coordinates, $q_i^{M\tilde m}$ and $\tilde q_{i'}^{M\tilde m'}$, which are therefore in a two-dimensional 
representation of the $SU(2)\times SU(2)$ group and are charged under the $U(1)$. However, since the orientifold projection identifies the gauge groups, 
as well as the $D2-D6$ fields, so too do the $q$ and $\tilde q$ combine. 
At this point, we can either describe this as taking $\tilde m,\tilde m'=1,...,4$ and combining them into $m=1,...,8$, or we can consider that both indices take eight values, but then $q$ is identified with $\tilde q$. Either way, the result is that the fields $q_i^{Mm},\tilde q_i^{
Mm}$, satisfy the reality condition
\be
q_i^{Mm}=J_{ij}\epsilon^{MN}(q^\dagger)^{mj}_{N}; \;\;\;
\tilde q_i^{Mm}=J_{ij}\epsilon^{MN}(\tilde q^\dagger)^{mj}_{N}.
\ee
In summary then, the field content of our flavoured ABJM model ends up being $Z^{A(ii')}, Z^{A[ii']}, q_i^{Mm},\tilde q_i^{Mm}$.
To write the superpotential for the theory, we first recall that the superpotential in the usual ABJM model (without orientifolding), can be written in terms of bifundamental fields 
$B_i,A_i$ and auxiliary adjoint fields $\phi_i$, as
\be
{\cal W}_{ABJM}=\frac{k}{8\pi}\Tr[\phi_1^2-\phi_2^2]+\Tr[B_i\phi_1A_i]+\Tr[A_i\phi_2B_i],
\ee
or, eliminating the auxiliary fields, as the quartic form
\be
\mathcal{W}_{ABJM}=\frac{2\pi}{k}\Tr(A_iB_iA_jB_j-B_iA_iB_jA_j)=\frac{2\pi}{k}\Tr(A_1B_1A_2B_2-B_1A_1B_2A_2).
\ee
Then the superpotential for our model is 
\bea
{\cal W}&\sim& \epsilon_{M'N'}(Z_{(ii')}+Z_{[ii']})^{MM'}J^{i'j}(Z_{(jj')}+Z_{[jj']})^{MN'}J^{j'k}\times\cr
&&\times \epsilon_{P'R'}
(Z_{(kk')}+Z_{[kk']})^{NP'}J^{k'l}(Z_{(ll')}+Z_{[ll']})^{NR'}J^{l'i}\cr
&&+q^{iNm}\Big[\epsilon_{M'N'}(Z_{(ii')}+Z_{[ii']})^{MM'}J^{i'j'}(Z_{(j'j)}+Z_{[j'j]})^{MN'}\Big]\tilde q^{jNm}.
\eea
As a check, we verify what happens under Higgsing. This corresponds to decomposing 
the $Z^{A(ii')}$ into the seven scalars $X^{(ii')I'}$ and an eigth scalar that gets
eaten by the gauge fields to become dynamical, with all other fields unchanged. The resulting model matches the $D2-$brane analysis above as it should, providing a consistency check of the construction. 

\subsection{Comparison With Previous Constructions}
Another construction for adding fundamentals to ABJM was found in \cite{Hikida:2009tp}. It corresponds to basically adding a probe $D6-$brane without 
considering the issues of the flux on a compact space or of exact conformality of the system (which is required in order to have a gravity dual with an 
$AdS_4$ factor). In four dimensions, one can do the same by considering a probe $D7-$brane in $AdS_5\times S^5$, wrapping a codimension-2 cycle in $S^5$
(see, e.g., \cite{Karch:2002sh}). But this contruction is not without subtleties.
First among these is that any such system will be afflicted with the aforementioned  problem with the flux, in that we need a negative sink of flux on a compact space, 
otherwise the flux lines will meet at a singular point away from the $D7-$brane.\footnote{Having a $D-$brane on a collapsable cycle as opposed to a 
point charge does not help from the point of view of charge, though it avoids tadpoles due to its instability \cite{Karch:2002sh}: 
consider a D1-brane wrapping an $S^1$ cycle inside $S^2$. The $D1-$brane can shrink until the cycle is wraps is very small and 
it looks almost pointlike, say around the South Pole of the $S^2$. But then we have the same problem with the electric charge: the flux lines will meet 
again at the North Pole, where therefore there should be a sink of negative charge of equal absolute value.} This would manifest itself in general
by uncancelled tadpoles in the field theory. A second problem is that the theory cannot be exactly conformal because the uncancelled flux will set a scale. 
It would only be so in a limited energy range which, in the probe approximation, can be considered large enough, hence the gravity 
dual cannot be purely $AdS_5$ times another factor. These problems are solved by the construction in \cite{Aharony:1998xz}. which introduces an $O7-$plane of 
charge -4 and four $D7-$branes to compensate at the same fixed point. Consequently, there is no uncancelled flux on the compact space, and the field theory is exactly 
conformal.

The same conclusion applies to our case. The construction of \cite{Hikida:2009tp} is only valid in the probe approximation, and can be considered to be obtained by separating 
the $O6-$planes and the other $D6-$branes, and moving them far away on the compact space from the $D6-$brane we retain. The construction corresponds to a $D6-$brane wrapping a codimension-3 cycle in $\mathbb{ CP}^3$, which is in fact the same $S^3/\mathbb{Z}_2$ 
defined above, but without any orientifolding. We have, instead, considered this cycle to be 
the fixed plane of an $O6$ orientifold plane, and have added four $D6-$branes there. In \cite{Hikida:2009tp}, the cycle was initially defined in a different way, but one
can easily show it is the same cycle. It was also proven that it corresponds to a supersymmetric brane configuration, at it should.

The superpotential for the model in \cite{Hikida:2009tp} is similar to ours. In ${\cal N}=1$ language, the ABJM model has bifundamental fields $(A_1,A_2)$ in the $(N,\bar N)$ representation and $(B_1,B_2)$ in the $(\bar N, N)$ representation, and auxiliary 
adjoints $\phi_1,\phi_2$. The fundamental fields we add are $q_1,q_2$ in the $(N,1)$ and $(1,N)$ representation, and $\tilde q_1,\tilde q_2$ in the $(\bar N,1)$ and $(1,\bar N)$ 
representation. The ABJM model superpotential is 
\be
{\cal W}_{ABJM}=\frac{k}{8\pi}\Tr[\phi_1^2-\phi_2^2]+\Tr[B_i\phi_1A_i]+\Tr[A_i\phi_2B_i],
\ee
and the flavor deformation is 
\be
{\cal W}_{flavor}=\Tr[\tilde q_1\phi_1q_1]+\Tr[\tilde q_2\phi_2 q_2].
\ee
Clearly then, after eliminating the auxiliary fields, the construction is similar to ours.

To conclude this section, let's recall the symmetries fo the model. The R-symmetry is an$SU(2)_R$, acting on $(A_i,\bar B_i)$, with an internal $SU(2)$, acting on the doublets $(A_1,A_2)$ and $(B_1,B_2)$. We can also, of course, introduce several flavors $N_f$, to produce a global flavor symmetry group $SO(N_f)$.

\subsection{Applicability to condensed matter}

To find possibile applications of these flavoured models to condensed matter physics, we need to understand the physics of $N=1$ truncations. To this end, the first point to note is that the orientifold model is still nonabelian for for $N=1$, since we have now a $USp(2)$ gauge theory. There appears nothing to be done about this, as it is just the result of the orientifold procedure. We can however choose a global $U(1)$ charge inside the $SO(8)$ carried by the $q$'s, and in 
this way get fundamental fields $q,\tilde q$ and their conjugates, coupling to the local gauge group with charges $+1$ and $-1$, and contributing $+1$
to the global charge (the analog of electric charge for the condensed matter model), as we wanted. As an added bonus, in this case the theory is conformal. 

For the probe model in \cite{Hikida:2009tp}, we can now consider $N=1$, and the $q,\tilde q$ fields are in the fundamental of the resulting $U(1)$ gauge 
group, with positive and negative charges. 
If we choose several flavors, with a $SO(N_f)$ symmetry group, we can again choose a $U(1)$ subset that corresponds to the global charge with $+1$ charge
contributions, with the important caveat of the potential problems discussed above.

\section{Conclusions}
In this article, we have analyzed several ways in which one can obtain an abelian theory out of the ABJM model, for the purpose of simulating condensed matter models
of interest. In particular, we have analyzed features of a model used for, among other things, the description of compressible Fermi surfaces in \cite{Huijse:2011hp}.
We have seen that simply setting $N=1$ in the ABJM model does not work, since we obtain a {\it free abelian theory}, with scalars which are not charged 
under the $U(1)$ Maxwell gauge group (after using the Higgsing procedure in $(2+1)-$dimensions to go from CS to Maxwell gauge fields).

Instead, one possibility that we found is to generalize the nontrivial abelian reduction in  \cite{Mohammed:2012rd,Mohammed:2012gi} to include fermions. 
In this way, we obtained an abelian theory with ${\cal N}=2$ supersymmetry and six global $U(1)$ charges, a combination of which can be taken to be somewhat similar
to the global electric charge in condensed matter models, in that the positive and negative charges of various fields are different from the positive 
and negative local charges. Another possibility that we found was to add fundamental fields to the ABJM model. One can construct a $D2$-$D6$-$O6$ system giving a conformal 
field theory with a gauge group $USp(2N)$ and an $SO(8)$ flavor group which simulates well the global electric charge of the condensed matter model. Our construction, while comparing well with existing literature, is, in addition, consistent with Gauss' law and exactly conformal, both of which we think make our model an excellent laboratory within which to realize the AdS/CMT correspondence in string theory. 

{\bf Acknowledgements}\\
We would like to thank Tadashi Takayanagi for useful discussions. HN would like to thank the University of Cape Town for hospitality 
during the time this project was started and completed.
The work of HN is supported in part by CNPq grant 301219/2010-9. JM acknowledges support from the National Research Foundation (NRF) of South Africa under the Incentive Funding for Rated Researchers and Thuthuka programs.

\bibliography{abfund}
\bibliographystyle{utphys}

\end{document}